\documentclass[12pt]{article}
\usepackage{amssymb}
\usepackage{amsmath, amsthm}
\newcommand{\be}{\begin{equation}}
\newcommand{\ee}{\end{equation}}
\begin{document}
\def\theequation{\arabic{section}.\arabic{equation}}
\begin{titlepage}
\title{The (pseudo)issue of the conformal frame revisited}
\author{Valerio Faraoni and Shahn Nadeau\\ \\
{\small \it Physics Department, Bishop's University}\\
{\small \it 2600 College Street,  Sherbrooke, Qu\`{e}bec, Canada 
J1M 0C8}
}
\date{} \maketitle
\thispagestyle{empty}
\vspace*{1truecm}
\begin{abstract}
The issue of the equivalence between Jordan and Einstein 
conformal frames in scalar-tensor gravity is revisited, 
with 
emphasis on implementing running units in the latter. The lack 
of affine parametrization for timelike worldlines and the 
cosmological constant problem in the Einstein frame are 
clarified, and a paradox in the literature about cosmological 
singularities appearing only in one frame is solved. While, 
classically, the two conformal frames are physically equivalent, 
they seem to be inequivalent at the quantum level.
  \end{abstract} \vspace*{1truecm} 

\begin{center}  {\bf Keywords:} scalar-tensor gravity, 
conformal transformations.
\end{center}
\begin{center} {\bf PACS:}   98.80.-k, 04.90.+e, 04.50.+h
\end{center}
\setcounter{page}{0}
\end{titlepage}

\def\theequation{\arabic{section}.\arabic{equation}}


\section{Introduction}

Conformal (or Weyl) transformations are widely used in 
scalar-tensor theories of gravity \cite{ST}, the theory of a 
scalar field 
coupled non-minimally
to the Ricci curvature $R$, and in modified gravity theories in 
which
terms non-linear in $R$ are added to the Einstein-Hilbert action 
(due
perhaps to quantum corrections \cite{Stelle}). The present 
acceleration of the universe discovered with the study of 
supernovae of type  Ia \cite{SN} calls either for an exotic form 
of dark energy (in 
Einstein
gravity or in scalar-tensor theories), or for modifications of gravity
described by terms non-linear in $R$ in the Lagrangian  
\cite{modifiedgravity}, or the addition of terms containing the 
invariants of the  Riemann tensor $R_{ab}R^{ab} $ and 
$R_{abcd}R^{abcd}$ \cite{PQ}. To 
fix the ideas  and the terminology, consider a scalar-tensor 
theory of gravity, 
described in the 
Jordan frame by the action \cite{footnote1} 
\begin{equation}
S=\int d^{4}x\; \sqrt{-g}\left[\frac{f(\phi)R}{2} 
-\frac{\omega(\phi)}{2}g^{ab}\nabla_{a}\phi\nabla_{b} 
\phi-V(\phi) 
\right]+\alpha_{m}\mathcal{L}^{(m)}\left[g_{ab},\psi_{m}\right]\;,
\label{eq:1.1}
\end{equation}
where $f>0$, $S^{(m)}=\int d^{4}x\;\sqrt{-g} 
\, \mathcal{L}^{(m)}$,
and  $\mathcal{L}^{(m)}$ is the Lagrangian density describing 
``ordinary''
matter (as opposed to the gravitational scalar field $\phi$, which
effectively plays the role of a form of non-conventional matter in
the field equations. Here $g_{ab}$ is the metric tensor with determinant
$g$, $f(\phi)$ and $\omega(\phi)$ are arbitrary coupling functions,
$\phi$ is the Brans-Dicke-like scalar field with potential 
$V(\phi)$,
and $\psi_{m}$ collectively denotes the matter fields. The Jordan
frame in which the theory (\ref{eq:1.1}) is formulated is the set
of dynamical variables $(g_{ab},\phi)$ describing the gravitational
field. The effective gravitational coupling is
\begin{equation}
G_{eff}=\frac{1}{8\pi f(\phi)}\;,\label{eq:1.1bis}
\end{equation}
as can be immediately deduced from inspection of the action 
(\ref{eq:1.1}).
However, in a Cavendish experiment the effective coupling is  
instead \cite{Nordvedt,Willbook}
\begin{equation}
G_{eff}^{(*)}=\frac{2\omega f+2(\frac{df}{d\phi})^{2}}{8\pi 
f\left[2\omega f+3(\frac{df}{d\phi})^{2}\right]}\;.
\end{equation}
 This expression can also be derived from
cosmological perturbation theory \cite{Boisseauetal00}. Note that
in the Jordan frame description the Lagrangian density $\mathcal{L}^{(m)}\left[g_{ab},\psi_{m}\right]$
only depends on the metric $g_{ab}$ and the ``ordinary'' matter
fields $\psi_{m}$. As a consequence, this matter is described by
the stress-energy tensor
\begin{equation}
T_{ab}^{(m)}=\frac{-2}{\sqrt{-g}}\frac{\delta S^{(m)}}{\delta g^{ab}}\;,\label{eq:1.2}\end{equation}
and the invariance of $S^{(m)}$ under diffeomorphisms leads to the
covariant conservation of $T_{ab}^{(m)}$ \cite{Wald}
\begin{equation}
\nabla^{b}T_{ab}^{(m)}=0\;.
\end{equation}
As a consequence, test particles in the Jordan frame follow 
geodesics, the weak equivalence principle \cite{Willbook} is 
satisfied, and the theory (\ref{eq:1.1}) is metric. In this 
frame the kinetic energy term of the scalar $\phi$, i.e., 
$- \omega(\phi)\nabla^{a}\phi\nabla_{a}\phi /2 $,
is non-canonical and has indefinite sign. The 
experimental constraint $\left|\omega(\phi_{0}) 
\right| >40000 $, where $\phi_{0}$ is the present
value of the scalar field, applies \cite{BertottiIessTortora}, 
unless a potential $V(\phi)$ gives the field a very short 
range.

Let us consider now the conformal transformation 
\begin{equation}
g_{ab} 
\longrightarrow \tilde{g}_{ab}= 
\Omega^{2}g_{ab} \;,\;\;\;\;\; 
\quad\Omega=\sqrt{f(\phi)}\;,\label{eq:1.4}
\end{equation}
and the scalar field redefinition
\begin{equation}
\phi \longrightarrow \tilde{\phi}=\int 
\frac{d\phi}{f(\phi)}\sqrt{f(\phi)+\frac{3}{2}\left(\frac{df}{d\phi} 
\right)^{2}}\;.\label{eq:1.5}
\end{equation}
This transformation brings the theory into the Einstein 
conformal frame, i.e., to the set of variables 
($\tilde{g}_{ab},\tilde{\phi}$)
in which the action (\ref{eq:1.1}) takes the form
\begin{equation}
S=\int d^{4}x\;\sqrt{-\tilde{g}}\left\{ \frac{\tilde{R}}{16\pi} 
-\frac{1}{2}\, 
\tilde{g}^{ab} \, 
\tilde{\nabla}_{a} 
\phi 
\tilde{\nabla}_{b}\phi 
-\tilde{U}(\tilde{\phi)} 
+(G\phi)^{-2}\mathcal{L}^{(m)}\left[\tilde{g}_{ab},\psi_{m}\right] 
\right\} \;,\label{eq:1.6}
\end{equation} 
where 
\begin{equation}
\tilde{U}(\tilde{\phi})=\frac{V\left[\phi(\tilde{\phi})\right]}{\Omega^{2}}\;,\label{eq:1.7}\end{equation}
and $\tilde{\nabla}_{a}$ is the covariant derivative operator of
the metric $\tilde{g}_{ab}$. Note that the ``new'' scalar 
field $\tilde{\phi}$ exhibits a
canonical kinetic energy term and it couples minimally to the 
Ricci
curvature $\tilde{R}$ of the ``new'' metric $\tilde{g}_{ab}$.
However, the action (\ref{eq:1.6}) does not describe simply general
relativity with an extra scalar field $\tilde{\phi}$, because $\tilde{\phi}$
couples explicitly to matter via the prefactor $\left[G\phi(\tilde{\phi})\right]^{-2}$
in front of the matter Lagrangian $\mathcal{L}^{(m)}$. The  
exceptions are forms of conformal matter which obey equations 
invariant under the conformal transformation 
(\ref{eq:1.4}), such as the 
Maxwell  field, a radiation fluid, or a scalar field $\psi$ 
conformally coupled to $R$ and with zero
or quartic potential. The explicit coupling
of $\tilde{\phi}$ to all forms of  non-conformal matter spoils 
the equivalence principle in the Einstein frame. In this frame 
all massive particles deviate from geodesics due to the 
force, proportional to
$\tilde{\nabla}_{a}\tilde{\phi}$,  exerted by $\tilde{\phi}$. By
contrast, zero mass particles still move along null geodesics. (This
can be realized by noting that the conformal transformation 
(\ref{eq:1.4})
does not change the conformally invariant Maxwell equations in four
dimensions, which reduce to geometric optics in the high frequency
limit.)

The issue has been raised of ``which conformal frame is physical'',
i.e., should one regard the Jordan frame metric $g_{ab}$, or the
Einstein frame metric $\tilde{g}_{ab}$ as physical? This issue has
been the subject of much debate and is still contentious due to incorrect
formulations of this question. In fact, the question is answered,
to a large extent, by Dicke's paper \cite{Dicke} which 
originally introduced  the conformal transformation for 
Brans-Dicke theory \cite{BransDicke},
the prototype of scalar-tensor gravity theories. The answer of 
Ref.~\cite{Dicke} is that the two frames
are equivalent, provided that the units of mass, length, time, and
quantities derived therefrom scale with appropriate powers of the
conformal factor $\Omega$ in the Einstein frame. However, Dicke's
treatment is valid only at the classical level, while in modern cosmology
and in gravitational theories alternative to Einstein gravity, quantum
fields in curved space play a significant role and the equivalence
of the conformal frames is not clear at all - indeed there are 
certain indications
that the equivalence breaks down at the quantum level. Of course,
nothing is known about this equivalence in quantum gravity due to
the lack of a definitive theory of quantum gravity.

In view of Dicke's paper, many authors consider the issue of which
conformal frame is physical a pseudo problem, and we agree with them
to a large extent, apart from the two problems mentioned above. However,
while the answer to the question of the physical equivalence of 
conformal frames may be clear in principle, its 
\emph{application} to practical situations
is a completely different matter. The scaling of units in the   
Einstein frame is usually forgotten or not taken into account, 
producing results that range from nonsensical to marginally 
incorrect, to correct but it is not easy to understand if the 
conformal transformation (\ref{eq:1.4})
and (\ref{eq:1.5}) is applied correctly. (It is worth noting that
Dicke himself applied the conformal transformation and the scaling
of units incorrectly in the simpler context of Einstein 
gravity \cite{DickePeebles}). Misinterpretations of the 
conformal 
transformation
abound in the literature and fuel the debate on the issue (or pseudo-issue)
of the conformal frame, while other authors consider the problem a
closed one and sharply state that the Einstein frame is physical while
the Jordan frame should not be considered at all. The argument for
this choice is the positivity of the kinetic energy and the existence
of a ground state, but this argument is usually not explored in detail
for the specific theories considered. The existing review papers on
the subject \cite{MagnanoSokolowski,FGN} fail to  
clarify this issue because they do not explicitly state the 
assumptions made.
It appears that they refer to a version of scalar-tensor gravity 
in the Einstein
frame in which the units of mass, length, time, etc. \emph{do not
scale} with powers of $\Omega$. This version of the theory has nothing
to do with the original Jordan frame and it is physically inequivalent
to it, but it has come to be implicitly accepted as a valid theory,
which adds to the confusion. It is our opinion that the issue deserves
some clarification and that the open problems (Cauchy problem, extension
to quantum matter) should be clearly formulated and addressed. In
this paper we state as clearly as possible what the problems are,
and we show how the divergence of opinions between different authors
is due to the fact that two physically different theories in the Einstein
frame (with or without scaling of units, respectively) are 
considered by different
authors without realizing, or explicitly stating, which one is the
version under examination. As a consequence,  much of the 
existing debate becomes meaningless, the two opposite viewpoints 
are both correct but they really refer to physically different 
theories (one of which not as well motivated as the other), 
while they are erroneously reported as pertaining to the same 
physical theory. From a more conservative point of view, 
instead, only the Einstein frame version of the theory 
incorporating scaling units is physically motivated. Even 
accepting this point of view, however, it is not always obvious 
how to incorporate this scaling of units in a calculation,
for example computing the spectrum of inflationary perturbations 
in scalar-tensor gravity, and this issue deserves some 
attention.

Many (perhaps most) researchers in gravitation and cosmology are unaware
of the importance of scaling units in the Einstein frame, which is
neglected. This issue is discussed in Sec.~2, where it is shown
that the scaling of units is related to the ``anomalous'' coupling
of the scalar $\tilde{\phi}$ to matter in the Einstein frame, and
to the subsequent violation of the equivalence principle. In 
Sec.~3 we examine some consequences of allowing or not the units 
of fundamental quantities to scale in the Einstein frame, and we 
resolve an apparent paradox in the literature regarding energy 
conditions and singularity theorems in the two conformal frames 
in Sec.~4. The cosmological constant problem and the 
Cauchy problem are discussed in Sec.~5, while Sec.~6 contains 
the conclusions.

\section{Conformal transformations, Jordan frame, and Einstein 
frame}
\setcounter{equation}{0}

Here we recall the basic properties of the conformal 
transformation to the Einstein frame, the transformation 
properties of various geometrical
quantities, and the conservation equations for the matter stress-energy
tensor. The  reader is referred to  
Refs.~\cite{Synge,Wald,MagnanoSokolowski,FGN} for further 
details.

Consider a spacetime ($M,g_{ab}$) where $M$ is a smooth 
manifold with dimension $n>1$ and $g_{ab}$ is a Lorentzian 
or Riemannian metric on $M$. The conformal  
transformation
\begin{equation}
g_{ab}\longrightarrow 
\tilde{g}_{ab} =\Omega^{2}g_{ab}\;,\label{eq:2.1}
\end{equation}
where $\Omega$ is a smooth, nowhere vanishing, function of the 
spacetime point is a point-dependent rescaling of the metric. It 
changes the length of timelike and spacelike intervals and 
vectors, but it preserves their timelike or spacelike character. 
Similarly, null intervals and null vectors according to the 
``old'' metric $g_{ab}$ remain null
according to the ``new'' metric $\tilde{g}_{ab}$. The light 
cones are not changed by the conformal transformation 
(\ref{eq:2.1}) and the spacetimes ($M,g_{ab}$) and 
($M,\tilde{g}_{ab}$) have the same causal structure; the 
converse is also  true \cite{Wald}. The inverse metric $g^{ab}$, 
the metric determinant $g$, and the Christoffel
symbols transform according to \cite{Synge,Wald}
\begin{equation}
\tilde{g}^{ab}=\Omega^{-2}g^{ab}\;,\qquad\tilde{g} 
=\Omega^{2n}g\;,\label{eq:2.2}
\end{equation}
\begin{equation}
\tilde{\Gamma}_{bc}^{a}=\Gamma_{bc}^{a}+\frac{1}{\Omega} 
\left(\delta_{b}^{a}\nabla_{c}\Omega+\delta_{c}^{a}\nabla_{b}
\Omega-g_{bc}\nabla^{a}\Omega\right)\;,\label{eq:2.3}
\end{equation}
while the Riemann and Ricci tensor obey
\begin{eqnarray}
\tilde{ {R_{abc}}^d} &= & {R_{abc}}^d +2\delta_{\left[ a 
\right.}^d 
\nabla_{b \left. \right]} \nabla_c \left( \ln \Omega \right)
-2g^{de}\, g_{c\left[ \right.} \nabla_{b \left. \right]} 
\nabla_e \left( \ln \Omega \right) \nonumber \\
&& \nonumber \\
 & + &  2\nabla_{\left[ a \right.} \left( \ln \Omega \right) 
\delta_{\left. b \right]}^d \nabla_c \left( \ln \Omega \right) 
-2\nabla_{\left[ a \right.} \left( \ln \Omega \right) g_{b 
\left. \right]c} \, g^{de}\nabla_e \left( \ln \Omega \right)  
\nonumber \\
&& \nonumber \\
&-& 2  g_{c\left[ a \right.} \delta_{\left. b \right]}^d \, 
g^{ef}  \nabla_e \left( \ln \Omega \right) \nabla_f \left( \ln 
\Omega  \right) \;,
 \label{eq:2.4} 
\end{eqnarray}
\begin{eqnarray}
 \tilde{R}_{ab}  &=& R_{ab}-\left( n-2\right) \nabla_a \nabla_b 
\left( \ln \Omega \right) -g_{ab}\, g^{ef} \nabla_f \nabla_e 
\left( \ln \Omega \right) \nonumber \\
&& \nonumber \\
& + & \left( n-2 \right) \nabla_a \left( \ln \Omega \right) 
\nabla_b \left( \ln \Omega \right) \nonumber \\
&& \nonumber \\
& - & \left( n-2 \right) g_{ab} \, g^{ef} \nabla_f \left( \ln 
\Omega \right) \nabla_e \left( \ln \Omega\right) \;.
\label{eq:2.5}
\end{eqnarray}
For the Ricci curvature,
\begin{equation}
\tilde{R}=\tilde{g}^{ab}\tilde{R}_{ab} = 
\frac{1}{\Omega^{2}}\left[R-2\left(n-1\right) 
\square\left(\ln\Omega\right)-\left(n-1\right) 
\left(n-2\right) \frac{g^{ab} \, \nabla_{a}\Omega 
\nabla_{b}\Omega}{\Omega^{2}}\right]\;.\label{eq:2.6}
\end{equation}
In $n=4$ dimensions it is
\begin{equation}
\tilde{R}=\frac{1}{\Omega^{2}}\left(R-\frac{6\square\Omega}{\Omega} 
\right)=\frac{1}{\Omega^{2}}\left[R-\frac{12\square\left( 
\sqrt{\Omega}\right)}{\sqrt{\Omega}}+3 \, 
\frac{g^{ab} 
\, \nabla_{a}\Omega\nabla_{b}\Omega}{\Omega^{2}}\right]\;.
\label{eq:2.7}
\end{equation}
The Weyl tensor $  { C_{abc}}^d$ with the last 
index raised is conformally invariant,
\begin{equation}
\tilde{  {C_{abc}}^d}= {C_{abc}}^d\;.\label{eq:2.8}
\end{equation}
However, the same tensor with the other indices raised or lowered
is not conformally invariant. Note that in the conformally rescaled
world the conformal factor $\Omega$ plays the role of a form of matter.
In fact, if the original metric is Ricci-flat ($R_{ab}=0$), the new
metric is not ($\tilde{R}_{ab}\neq0$).

If the Weyl tensor of $g_{ab}$ vanishes, also the Weyl tensor of
$\tilde{g}_{ab}$ in the conformally related frame vanishes (and vice-versa),
conformally flat metrics are  mapped into conformally flat 
metrics.

Let us consider covariant conservation for the matter 
energy-momentum tensor $T_{ab}^{(m)}$. In the Jordan frame it is  
\begin{equation}
\nabla^{b} T_{ab}^{(m)}=0\;;\label{eq:2.9}
\end{equation}
this equation is not conformally invariant and $T_{ab}$ scales 
as \cite{Wald}
\begin{equation}
\tilde{T}_{(m)}^{ab}=\Omega^{s} 
\, T_{(m)}^{ab}\;, \;\;\;\; \qquad\tilde{T}_{ab}^{(m)}
=\Omega^{s+4} \, T_{ab}^{(m)}\;,\label{eq:2.10}
\end{equation}
where $ s $ is a appropriate conformal weight. As a consequence, 
the conservation equation in the conformally rescaled world is
\begin{equation}
\tilde{\nabla}_{a} \left(\Omega^{s} \,  T_{(m)}^{ab}\right)= 
\Omega^{s} \, 
\nabla_{a}T_{(m)}^{ab} 
+ \left(s+6\right)\Omega^{s-1}T_{(m)}^{ab}\nabla_{a}
\Omega-\Omega^{s-1}g^{ab } 
\, T^{(m)}\nabla_{a}\Omega\;,\label{eq:2.11}
\end{equation}
in four spacetime dimensions \cite{Wald}. By conveniently 
choosing $ s=-6 $ one obtains
\begin{equation}
\tilde{\nabla}_{a}\tilde{T}_{(m)}^{ab}= 
-\tilde{T}^{(m)} \, \tilde{g}^{ab} \tilde{\nabla}_{a} 
\left(\ln\Omega\right) \label{eq:2.12}
\end{equation} 
and
\begin{equation}
\tilde{T}^{(m)}\equiv\tilde{g}^{ab} 
\tilde{T}_{ab}^{(m)}=\Omega^{-4} 
\, T^{(m)}\;.\label{eq:2.13}
\end{equation}
Hence, in the new conformal frame, the stress-energy tensor 
$\tilde{T}_{ab}^{(m)}$ is not covariantly conserved unless it 
describes conformally invariant matter with vanishing trace 
$T^{(m)}=0$, in which case also $ \tilde{T}^{(m)}=0 $
and $\tilde{\nabla}^{b}\tilde{T}_{ab}^{(m)}=0 $.

It is  well known that null geodesics of the Jordan metric 
$g_{ab}$ are  mapped into null geodesics of the Einstein frame 
metric $\tilde{g}_{ab}$ \cite{Wald}. Timelike geodesics will be 
considered in the next 
section.

\section{Jordan frame, Einstein frame with running units, and 
Einstein frame with fixed units}
\setcounter{equation}{0}

In this section only classical physics of spacetime 
and matter is considered. Quantum matter  will be discussed in 
Sec.~6.

A viewpoint shared by many authors (see  
\cite{MagnanoSokolowski,FGN} for references) states that the 
Einstein and Jordan conformal frames
are physically equivalent.  This viewpoint is generally correct  
as shown below, and it is in open conflict with the viewpoint 
that the Jordan frame should be abandoned in favour of the 
Einstein frame because of the presence of negative energy.

\subsection{Einstein frame with running units}

The argument of the physical equivalence between the Jordan and Einstein
frames dates back to Dicke's 1962 paper introducing the conformal
transformation technique for Brans-Dicke theory \cite{Dicke}, a paper
often forgotten or misread. The basic idea is that the two  
conformal frames are physically equivalent \emph{provided} that 
in the  Einstein
frame the units of time, length, mass, and derived quantities are
allowed  to scale with appropriate powers of the conformal 
factor $\Omega$. Physics must be invariant under a choice of
the units --- this includes not only transformations of units by 
factors
which are the same everywhere in spacetime (``rigid'' changes
of units or ``dilatations''), but also changes of units that depend
on the spacetime point. A rescaling of the units of length and time
(and, on dimensional grounds, also of mass) is a conformal transformation.
Since physics is invariant under a change of units, it is invariant
under a conformal transformation provided that the units of length,
time, and mass $l_{u}$, $t_{u}$, and $m_{u}$ are scaled. The novelty
of Dicke's approach consists in allowing these units to be rescaled
by different factors at different spacetime points, with the change
in each unit being a smooth, nowhere vanishing, function of the spacetime
point. Instead of a system of units rigidly attached to the spacetime
manifold, the Einstein frame contains a system of units that 
change
with the spacetime location. If one accepts this point of view, the
symmetry group of classical physics is enlarged to include conformal
transformations \emph{with the associated rescaling of units}. 

It is shown in Ref.~\cite{Dicke} that $g_{ab}$ scales with the 
dimensions of a time squared, and since  
$ \tilde{g}_{ab}=\Omega^{2}g_{ab}$, it follows that times and 
lengths scale with $\Omega$, so that 
\begin{equation}
dt  \rightarrow \tilde{dt}=\Omega \, dt\;, \;\;\;\;\; dx^{i}  
\rightarrow  \tilde{dx^{i}}=\Omega \,
dx^{i}\qquad(i=1,2,3)\;,\label{eq:3.4}
\end{equation}
while for masses
\begin{equation}
m\rightarrow\tilde{m}=\Omega^{-1}m\;,\label{eq:3.5}\end{equation}
on dimensional grounds.

Since the speed of light in vacuum $c$ is a ratio of space and time,
it is invariant and local Lorentz invariance is preserved. The Planck
constant, which has dimensions $\left[h\right]=\left[ML^{2}T^{-1}\right]$
is left unchanged, while energy with dimensions $\left[Mc^{2}\right]$
scales like a mass. In the Jordan frame of scalar-tensor gravity 
the effective coupling (\ref{eq:1.1bis}) varies, while $h$, $c$, 
the masses of elementary particles, and the coupling constants 
of physics are  true
constants, together with the units. The weak equivalence principle
holds and the theory is metric. On the contrary, in the Einstein frame
the gravitational coupling $G$ is constant and so are $h$ and $c$,
while the masses of elementary particles and the coupling ``constants''
of non-gravitational physics vary with time together with the units
of time, length, and mass $\tilde{t}_{u}$, $\tilde{l}_{u}$, and
$\tilde{m}_{u}$. In Dicke's viewpoint the Jordan and Einstein frames
are merely two equivalent representations of the same physics. One
can consider, for example, the proton mass, which has a constant value
$m_{p}$ in the Jordan frame. In the Einstein frame the proton mass
depends on $\Omega$ (or $\phi$) and is $\tilde{m}_{p}=\Omega^{-1}m_{p}$.
However, in an experiment one measures the  \emph{ratio} 
$ \tilde{m}_{p} / \tilde{m}_u $
between the proton mass and an arbitrarily chosen mass unit 
$\tilde{m}_{u}$. Hence, in the Einstein frame it is not 
$\tilde{m}_{p}$ that matters, but the ratio
\begin{equation}
\frac{\tilde{m}_{p}}{\tilde{m}_{u}} 
=\frac{\Omega^{-1}m_{p}}{\Omega^{-1}m_{u}}= 
\frac{m_{p}}{m_{u}}    \label{eq:3.6}
\end{equation}
(in the Jordan frame as  well, it is only $ m_{p} / m_u $ 
that is measured). A measurement of the proton mass with respect 
to the chosen mass unit therefore yields the same value in the 
Jordan and the Einstein frame. No preferred frame is selected by 
such a measurement. The outcome of an experiment is the same 
when analyzed in the Jordan or the Einstein frame. In this 
context, the problem of which frame is physical is void of 
content.

In the Einstein frame not only the masses of elementary 
particles and the mass units, but also the coupling constants of 
non-gravitational physics vary with $\phi$. To understand this 
variation it is useful to consider the following example, to 
which we  will return later \cite{footnote2}.

\subsubsection{Examples}

Consider as an example Brans-Dicke theory \cite{BransDicke} 
with a  massive Klein-Gordon field $\psi$ as the only form of 
matter, as described in the Jordan frame by the action
\begin{eqnarray}
S &=& S^{(BD)}+S^{(KG)}=\int d^4 x \sqrt{-g} \left( {\cal 
L}^{(BD)}+\alpha_{KG}\, {\cal L}^{(KG)} \right) \nonumber \\
&& \nonumber \\
&=& 
\int d^4 x \sqrt{-g} \left( \phi R -\frac{\omega}{\phi} \, 
g^{ab} \nabla_a\phi\nabla_b\phi \right) -\frac{\alpha_{KG}}{2} 
\int d^4 x \sqrt{-g} \left( g^{ab}\nabla_a\psi\nabla_b\psi 
+m^2\psi^2 \right)\;, \nonumber \\
&& \label{eq:3.7}
\end{eqnarray}
where $\alpha_{KG}=16\pi G$ is the Klein-Gordon coupling 
constant. The conformal transformation (\ref{eq:1.4}) and the 
scalar field
redefinition (\ref{eq:1.5}) yield
\begin{equation}
\sqrt{-g}\left(\mathcal{L}^{(BD)} +\alpha_{KG}\mathcal{L}^{(KG)} 
\right)=\sqrt{-\tilde{g}}\left\{  \tilde{\mathcal{L}}^{(GR)} 
+\tilde{\alpha}_{KG}(\phi) \, \tilde{\mathcal{L}}^{(KG)} 
\left[\tilde{g}_{ab},\psi\right]\right\} \;,\label{eq:3.8}
\end{equation}
where $ \tilde{\mathcal{L}}^{(GR)} 
= 
\tilde{R}-\frac{1}{2}\tilde{g}^{ab} 
\tilde{\nabla}_{a}\tilde{\phi}\tilde{\nabla}_{b}\tilde{\phi} $
is the Einstein-Hilbert Lagrangian  density with a canonical 
scalar field $\tilde{\phi}$,
\begin{equation}
\tilde{\alpha}_{KG}(\tilde{\phi})=\Omega^{-2}\alpha_{KG}= 32\pi 
G \exp\left( -8\sqrt{\frac{\pi 
G}{2\omega+3}} \, \tilde{\phi} \right) \;,\label{eq:3.10}
\end{equation}
\begin{equation}
\tilde{\mathcal{L}}^{(KG)}\left[\tilde{g}_{ab} ,\phi\right] 
= \frac{1}{2} \, \tilde{g}^{ab} 
\, 
\tilde{\nabla}_{a}\psi\tilde{\nabla}_{b}\psi+\frac{\tilde{m}^{2}}{2}\psi^{2}\;,\label{eq:3.10b}\end{equation}
and
\begin{equation}
\tilde{m}(\tilde{\phi})=\frac{m}{\Omega}= m 
\exp\left( -4\sqrt{\frac{\pi 
G}{2\omega+3}} \, \tilde{\phi} \right) \;,\label{eq:3.11}
\end{equation}
in accordance with eq.~(\ref{eq:3.5}). In the Einstein frame 
the mass $\tilde{m}$ of the Klein-Gordon field $\psi$ and its 
coupling constant $\tilde{\alpha}_{KG}$ acquire a dependence 
from the Brans-Dicke scalar. This holds  true for all forms of 
matter except  conformally invariant matter, which satisfies 
conformally invariant equations. As an example of such matter, 
consider the Maxwell field in four spacetime
dimensions, described by the matter action
\begin{eqnarray}
S^{(em)} & = & \int 
d^{4}x\;\sqrt{-g}\, \alpha_{em} \, \mathcal{L}^{(em)}\nonumber 
\\ 
& =  & -\int d^{4}x\;\sqrt{-g} \, F_{ab}F^{ab}\;,\label{eq:3.12}
\end{eqnarray}
with $\alpha_{em}=4$ and $\mathcal{L}^{(em)}=-\frac{1}{4}F_{ab}F^{ab}$,
where $F_{ab}$ is the antisymmetric Maxwell tensor. The conformal
invariance can be directly verified by computing
\begin{eqnarray}
\sqrt{-g}\, \mathcal{L}^{(em)} & = & -\frac{1}{4}\sqrt{-g}\; 
g^{ac}g^{bd}F_{ab}F_{cd}\nonumber \\
 & = & -\frac{1}{4}\left(\Omega^{-4}\sqrt{-\tilde{g}}\right)\left(\Omega^{2}\tilde{g}^{ac}\right)\left(\Omega^{2}\tilde{g}^{bd}\right)F_{ab}F_{cd}\nonumber \\
 & = & -\frac{1}{4}\sqrt{-\tilde{g}}\;\tilde{g}^{ac}\tilde{g}^{bd}\tilde{F}_{ab}\tilde{F}_{cd}\;,\label{eq:3.13}\end{eqnarray}
where $\tilde{F}_{ab}=F_{ab}$.

\subsubsection{Terminology}

Implementing the idea that physics should be conformally invariant
when units are rescaled leads to conflict with current terminology.
Consider, for example, a conformally coupled Klein-Gordon field $\psi$
with a non-zero mass, obeying the equation
\begin{equation}
\square\psi-\frac{R}{6}\psi-m^{2}\psi=0\;.\label{eq:3.14}
\end{equation}
According to standard terminology, the introduction of the mass $m$
breaks the conformal invariance that is present when $m=0$. One can,
however, generalize the notion of conformal invariance by allowing
the mass to vary with the scalar $\phi$. Upon the use of the relation
\begin{equation}
g^{ab} \nabla_{a}\nabla_{b}\psi-\frac{R}{6}\psi 
= 
\Omega^{3}\left[\tilde{g}^{ab}\tilde{\nabla}_{a} 
\tilde{\nabla}_{b}\tilde{\psi}-\frac{\tilde{R}}{6} 
\tilde{\psi}\right]\;,\label{eq:3.15}
\end{equation}
where $\tilde{\psi}\equiv\Omega^{-1}\psi$, one obtains from 
eq.~(\ref{eq:3.14})
\begin{equation}
\tilde{\square}\tilde{\psi}-\frac{\tilde{R}}{6}\tilde{\psi} 
-\tilde{m}^{2}\tilde{\psi}=0\;,\label{eq:3.16}
\end{equation}
where now $\tilde{m}\equiv\Omega^{-1}m$, in agreement with 
eq.~(\ref{eq:3.5}).
Hence, the Klein-Gordon equation (\ref{eq:3.14}) is invariant in
form if the current definition of conformal transformation is enlarged
to include the notion that masses scale with eq.~(\ref{eq:3.5}).
However, eq.~(\ref{eq:3.14}) is not conformally invariant 
according
to standard terminology.

\subsection{The equation of motion of  massive particles in the 
Einstein frame
with running units}

In the Einstein frame the equation of timelike geodesics receives
corrections and, as a result,  massive particles do not follow 
geodesics.
First, we want to find the transformation property of the four-velocity
$u^{a}= dx^a / d\lambda $ of a  massive particle, where 
$\lambda$
is a parameter along the geodesic, which is usually not discussed
in the literature. The Jordan frame normalization is $u^{a}u_{a}=-1$;
by assuming that $\tilde{u}_{a}=\Omega^{w}u_{a}$, where $w$ is an
appropriate conformal weight, and by imposing the Einstein frame  
normalization $ \tilde{g}^{ab}\tilde{u}_{a} \tilde{u}_{b}=-1$, 
one obtains $w=-1$,
or
\begin{equation}
\tilde{u}^{a}=\Omega^{-1} \, 
u^{a}\;,\;\;\;\;\qquad\tilde{u}_{a}=\Omega \, 
u_{a}\;.\label{eq:3.s1}\end{equation}
These relations can be used to find the relation between the parameters
$\lambda$ and $\tilde{\lambda}$ along the geodesic in the two conformal
frames. Since $u^{a}= dx^a / d\lambda $, 
$ \tilde{u}^a= d\tilde{x}^a / d\tilde{\lambda} $,
and lengths scale as $d\tilde{x}^{a}=\Omega dx^{a}$, by setting $d\tilde{\lambda}=\Omega^{\alpha}d\lambda$
one obtains $\tilde{u}^{a}=\Omega^{1-\alpha}u^{a}$ which, compared
with eq.~(\ref{eq:3.s1}) yields $\alpha=2$, or
\begin{equation}
d\tilde{\lambda}=\Omega^{2}d\lambda\;,\label{eq:3.s.2}\end{equation}
which agrees with eq.~(D.6) of Ref.~\cite{Wald}. This relation 
can
also be obtained from  the fact that, in terms of proper times 
$d\tau$
and $d\tilde{\tau}$, we have 
\begin{equation}
d\tilde{s}^{2}=-d\tilde{\tau}^{2}= 
\tilde{g}_{00}d\tilde{t}^{2}= 
\left( \Omega^{2}g_{00} 
\right) \left( \Omega^{2}dt^{2}\right)=-\Omega^{4}d\tau^{2}=
\Omega^{4}ds^{2}\;,
\end{equation}
which yields again $d\tilde{s}=\Omega^{2}ds$ for the parameter along
the timelike curve.

We are now ready to write the equation of motion of massive particles
in the Einstein frame. Under the conformal transformation (\ref{eq:2.1})
the Jordan frame geodesic equation $u^{a}\nabla_{a}u^{b}=0$ is  
mapped to \cite{Wald}
\begin{equation}
u^{a}\tilde{\nabla}_{a}u^{b}= 
2u^{b} \, \frac{u^{c} \nabla_{c}\Omega}{\Omega}  
+\frac{g^{bd} \tilde{\nabla}_{d}\Omega}{\Omega}\;.
\label{eq:3.s.3}
\end{equation}
By rewriting this equation in terms of tilded quantities we 
have
\begin{equation}
\tilde{u}^{a} \tilde{\nabla}_{a}\tilde{u}^{b} 
=\left(\frac{\tilde{u}^{c}\tilde{\nabla}_{c} 
\Omega}{\Omega}\right)\tilde{u}^{b}+ 
\frac{\tilde{g}^{bd} \tilde{\nabla}_{d}\Omega}{\Omega}\;.
\label{eq:3.s.4}
\end{equation}
The first term on the right hand side appears because the 
equation is not expressed using an affine parameter, while the 
second term proportional to the  gradient 
$\tilde{\nabla}_{a}\left(\ln\Omega\right)$
describes the direct  coupling of the field $\phi$ to 
non-conformal matter in the Einstein  frame; it has been likened 
to a fifth force violating the equivalence principle and making 
scalar-tensor theory in the Einstein
frame non-metric. It is impossible to achieve an affine 
parametrization
of this timelike curve and thus remove the first term on the right
hand side of  eq.~(\ref{eq:3.s.4}). In fact, if this could be 
achieved, the result would be incompatible with the 
normalization $\tilde{u}^{a}\tilde{u}_{a}=-1$.
To prove this statement, note that the normalization implies that
the four-acceleration $\tilde{a}^{b}\equiv\tilde{u}^{a}\tilde{\nabla}_{a}\tilde{u}^{b}$
is orthogonal to the four-velocity $\left(\tilde{u}^{b}\tilde{a}_{b}=0\right)$,
a well known fact  \cite{LandauLifschitz,Wald}. Then, 
\begin{equation}
0=\tilde{u}^{b}\tilde{a}_{b}\equiv\tilde{u}^{b}\tilde{u}^{a}\tilde{\nabla}_{a}\tilde{u}_{b}=\tilde{u}^{b}\tilde{\nabla}_{b}\Omega\;,
\end{equation}
implying that the gradient of the conformal factor must be orthogonal
to $\tilde{u}^{a}$ for \emph{any} possible  choice of 
$\tilde{u}^{a}$:
this is clearly absurd. For example, in scalar-tensor  cosmology where $\Omega=\Omega(t)$,
$t$ being the comoving time of a 
Friedmann-Lemaitre-Robertson-Walker metric (FLRW),
and by choosing $\tilde{u}^{a}$ as the four-velocity of comoving
observers, it follows that $ \partial\Omega / \partial t=0 $,
which is impossible 
\cite{footnote3}.  Therefore, the term 
$\left[ \tilde{u}^{c}\tilde{\nabla}_{c} 
\left(\ln\Omega\right)\right]\tilde{u}^{b} $
in eq.~(\ref{eq:3.s.4}) can not be eliminated or, in other 
words,
affine parametrization can not be achieved. 
Equation~(\ref{eq:3.s.4})
can be rewritten using eq.~(\ref{eq:3.5}) as
\begin{equation}
\tilde{u}^{a} 
\tilde{\nabla}_{a} \tilde{u}^{b}= 
-\left( 
\frac{ \tilde{u}^{c} \tilde{\nabla}_{c}\tilde{m}}{\tilde{m}} 
\right)\tilde{u}^{b}-\frac{ \tilde{g}^{bd} \,\tilde{\nabla}_{d} 
\tilde{m}}{\tilde{m}}\;.\label{eq:add1}
\end{equation}
Equation~(\ref{eq:add1}) suggests the interpretation that 
massive particles
deviate from geodesics because their mass is a function of the spacetime
point, and this deviation is proportional to the mass gradient. The
impossibility of using an affine parametrization is then traced back
to the impossibility of eliminating the variation $\tilde{u}^{c}\tilde{\nabla}_{c}\tilde{m}$
of the mass $\tilde{m}$ along the direction of motion of the particle.
By introducing the three-dimensional metric on the 3-space 
orthogonal to the four-velocity
$\tilde{u}^{a}$ of the particle,
\begin{equation}
\tilde{h}_{ab}\equiv\tilde{g}_{ab}+\tilde{u}_{a}\tilde{u}_{b}\;,
\end{equation}
where $h_{\: b}^{a}$ is the projection operator on the 3-space of
the observer $u^{a}$ (i.e., $ {h^a}_b \, u^{b}={h_a}^b   
\, u^{a}=0$), eq.~(\ref{eq:add1}) is rewritten as
\begin{equation}
\tilde{u}^{a}\tilde{\nabla}_{a}\tilde{u}^{b} 
=-\frac{\tilde{h}^{bd} \tilde{\nabla}_{d}\tilde{m}}{\tilde{m}} 
\;,\label{eq:add2}
\end{equation}
which shows explicitly that the correction to the equation of 
motion is  given entirely by the variation of the particle mass 
$\tilde{m}$  in the 3-space of an observer moving with the 
particle. Removing  the term  $  -\left(  
\tilde{u}^{c} \tilde{\nabla}_{c}\tilde{m} / \tilde{m} \right) 
\tilde{u}^b $ 
from eq.~(\ref{eq:add1}) by means of affinely parametrizing the 
curve would mean introducing corrections to the right hand side 
of  eq.~(\ref{eq:add2}) which are proportional to the derivative 
of $\tilde{m}$ in the direction of motion $\tilde{u}^{a}$, and 
this is impossible. It would mean that the right hand side of 
eq.~(\ref{eq:add1}) could not be  written explicitly as a 
purely spatial vector, as is instead done  in 
eq.~(\ref{eq:add2}), and therefore it could not be the 
four-acceleration $ a^{b}=u^{a}\nabla_{a}u^{b}$, which satisfies  
$ u^{a}a_{a}=0 $.

Equation (\ref{eq:add2}) has consequences for cosmology. In the 
FLRW metric
\begin{equation}
ds^{2}=-dt^{2}+a^{2}(t)\left[ 
\frac{dr^{2}}{1-Kr^{2}}+r^{2}\left(d\theta^{2} +\sin^{2}\theta 
d\varphi^{2}\right)\right]\;,\label{eq:add3}
\end{equation}
let $u^{a}$ be the four-velocity of comoving observers. Since in
FLRW scalar-tensor cosmology the scalar 
field depends only on the comoving
time in order to preserve spatial homogeneity, it is $\phi=\phi(t)$,
$\Omega=\Omega(t)$, and $\tilde{m}=\tilde{m}(t)$, which implies
that the spatial gradient $\tilde{h}^{bd}\tilde{\nabla}_{d}\tilde{m}$
vanishes identically, and the equation of motion of comoving observers,
which is the equation of timelike geodesics when a dust fluid with
pressure $P=0$ fills the universe, receives no correction in the
Einstein frame. Similarly, when there is pressure, the  timelike 
geodesic equation gets corrected by an extra term  in $P$, but 
no ``fifth force''
corrections  $-\tilde{h}^{bd} 
\tilde{\nabla}_{d}\left(\ln\tilde{m}\right)$ appear. The 
equivalence 
between Jordan and Einstein frames with respect to redshift, 
Boltzmann equation, and particle physics reaction rates in the 
early universe is discussed in Ref.~\cite{Catenaetal}.

The trajectories of particles with zero mass $\tilde{m}=m=0$ do 
not receive corrections when going to the Einstein frame.

\subsection{Einstein frame with fixed units}

By now it is clear that if one performs the conformal 
transformation (\ref{eq:1.4}) and (\ref{eq:1.5}) but does not 
allow the  units  of length, time, and mass to scale with 
$\Omega$ in the Einstein
frame (``fixed units''), one obtains a different physical
theory altogether. In this case, the conformal transformation is merely
a mathematical device relating the two conformal frames, and the Jordan
and Einstein frame are physically inequivalent. If the Jordan frame
and the Einstein frame \emph{with fixed units} were physically equivalent,
it would mean that the entire realm of (classical) physics is 
conformally
invariant, according to current terminology. But, to quote an 
example,
the Klein-Gordon field obeying eq.~(\ref{eq:3.14}) with $m\neq0$
is not conformally invariant in this sense. As another example, consider
conformally related metrics which are physically inequivalent such
as the Minkowski metric $\eta_{ab}$ and the  
FLRW metric given by the line element
\begin{equation}
ds^{2}=g_{ab} \, dx^{a}dx^{b}=a^{2}(\eta) 
\left(-d\eta^{2}+dx^{2}+dy^{2}+dz^{2}\right)\;,\label{eq:3.17}
\end{equation}
where $\eta$ is conformal time and $g_{ab}$ is manifestly 
conformally flat: $g_{ab}=\Omega^{2}\eta_{ab}$ with 
$\Omega(\eta)=a$. When $ da / d\eta >0$,
$g_{ab}$ describes an expanding universe with spacetime curvature,
cosmological redshift, possibly a Big Bang and/or other singularities,
and matter. By contrast, the conformally related metric $\eta_{ab}$
can not be associated to any of these spacetime features. The 
two
metrics $g_{ab}$ and $\eta_{ab}$ are physically equivalent only
when the fundamental units are allowed to scale with $a(\eta)$ in
the spirit of Refs.~\cite{Dicke,DickePeebles}. Then, the 
universe (\ref{eq:3.17}) appears flat when the units of time and 
length scale as $ dt = a (\eta) d\eta$,
$d \tilde{x}^{i}=a(\eta)dx^{i}$, giving (see \cite{FGN} for 
references)
\begin{equation}
ds^{2}=-d\tilde{t}^{2}+d\tilde{x}^{2}+d\tilde{y}^{2}+d\tilde{z}^{2}\;.\label{eq:3.18}\end{equation}
The recurring debate on the issue of which conformal frame  is 
physical arises from  the fact that many authors refer to the 
Einstein 
frame
by keeping the fundamental units fixed in this frame. The result is
a theory, which we shall call ``Einstein frame with fixed units''
version, which is physically inequivalent to the Jordan frame version
of scalar-tensor   gravity. These same authors often claim that 
the Jordan frame
version and the ``Einstein frame with fixed units version'' are
equivalent, forgetting about the scaling of units and Dicke's paper.
The Einstein frame-with-fixed-units version does not share the physical
motivations that lead to its Jordan frame cousin. It can even be said
that the former arises from a mistake, but given the number of works
devoted to this ``wrong'' theory, we are perhaps facing an 
(unintentional) new theory of gravity. We leave to the reader 
the judgment
of whether there is enough physical motivation to pursue 
``Einstein
frame with fixed units'' versions of gravitational theories, and
we content ourselves  to 
clarify the issue \cite{footnote4}.

Let us return for a moment to the Einstein frame with running units:
in this frame $\tilde{m}(\phi)=\Omega^{-1}m$ and the ratio of the
mass of a particle to the variable mass unit is constant,
\begin{equation}
\frac{\tilde{m}(\phi)}{\tilde{m}_{u}(\phi)}
=\mbox{constant}\;.
\end{equation}
This implies that
\begin{equation}
\frac{\tilde{\nabla}_{c}\tilde{m}}{\tilde{m}}= 
\frac{\tilde{\nabla}_{c}\tilde{m}_{u}}{\tilde{m}_{u}}\;,
\end{equation}
and therefore the equation of motion of massive particles (\ref{eq:add2})
in the Einstein frame with running units can be  written as
\begin{equation}
\tilde{u}^{a} \tilde{\nabla}_{a} \tilde{u}^{b} 
= -\frac{ \tilde{h}^{bd} \tilde{\nabla}_d  
\tilde{m}_{u}}{ \tilde{m}_{u}}\;;
\end{equation}
in other words, the correction to the equation of timelike geodesics
and the violation of the equivalence principle can be seen as arising
completely from the variation of the mass unit. Therefore, in the
Einstein frame with fixed units this correction vanishes and the equivalence
principle is satisfied unless, of course, one reintroduces these violations
by hand into the theory, but the latter now bears no relation to the
original Jordan frame one.

The running of fundamental units can also be seen as  the fact 
that
there is an anomalous coupling of $\tilde{\phi}$ to the matter sector
in the Einstein frame with running units. This  will be clear at 
the
end of this section. We now want to make contact with a different
notation that appeared recently in Ref.~\cite{Flanagan}. The 
author E. Flanagan parametrizes different conformal frames of a 
scalar-tensor   
theory using three
different functions of the scalar field $A(\phi)$, $B(\phi)$, and
$\alpha(\phi)$. The action is written as
\begin{equation}
S=\int d^{4}x\;\sqrt{-g}\left[\frac{A(\phi)}{16\pi G}R-\frac{B(\phi)}{2}g^{ab}\nabla_{a}\phi\nabla_{b}\phi-V(\phi)\right]+S^{(m)}\left[e^{2\alpha(\phi)}g_{ab},\psi^{(m)}\right]\;.\label{eq:3.19}\end{equation}
A conformal transformation is described by 
\begin{equation}
g_{ab} \longrightarrow\tilde{g}_{ab}=  
e^{-2\gamma(\phi)}g_{ab}\;,  \label{eq:3add1}
\end{equation}
\begin{equation}
\Phi\longrightarrow\tilde{\Phi}= 
h^{-1}(\phi)\;, 
\;\;\;\;\; 
\mbox{or} \;\;\;\; \qquad\Phi=h(\tilde{\phi})\;,\label{eq:3add2}
\end{equation}
where $\gamma$ and $h$ are regular functions with $h'>0$. The 
action can be rewritten as
\begin{equation}
S=\int d^{4}x\;\sqrt{-g}\left[\frac{\tilde{A}(\tilde{\phi})}{16\pi G}\tilde{R}-\frac{\tilde{B}(\tilde{\phi)}}{2}\tilde{g}^{ab}\tilde{\nabla}_{a}\tilde{\phi}\tilde{\nabla}_{b}\tilde{\phi}-\tilde{V}(\tilde{\phi})\right]+S^{(m)}\left[e^{2\tilde{\alpha}(\tilde{\phi})}\tilde{g}_{ab},\psi^{(m)}\right]\;,\label{eq:3add3}\end{equation}
where
\begin{eqnarray}
\tilde{\alpha}(\tilde{\phi}) 
&= &\alpha\left[h(\tilde{\phi})\right] 
+\gamma(\tilde{\phi})\;,\label{eq:3add4} \\
&&\nonumber \\
\tilde{V}(\tilde{\phi})&= 
& 
e^{4\gamma(\tilde{\phi})} 
V\left[h(\tilde{\phi})\right]\;,\label{eq:3add5} \\
&&\nonumber \\
\tilde{A}(\tilde{\phi})&=&e^{2\gamma(\tilde{\phi})} 
A\left[h(\tilde{\phi})\right]\;,\label{eq:3add6} \\
&&\nonumber \\
\tilde{B}(\tilde{\phi})&=& e^{2\gamma(\tilde{\phi})}\left\{ 
h'(\tilde{\phi})B\left[h(\tilde{\phi})\right]
-\frac{3}{4\pi G}h'(\tilde{\phi})\gamma'(\tilde{\phi}) 
A'\left[h(\tilde{\phi})\right]-\frac{3}{4\pi G} 
\left[\gamma'(\tilde{\phi})\right]^{2}A\left[h(\tilde{\phi}) 
\right]\right\} \;.\label{eq:3add7} \nonumber \\
&&\nonumber \\
&&
\end{eqnarray}
$\bullet$~~~In these notations the \emph{Jordan frame} 
corresponds to the  choice
\begin{equation}
\alpha=0 \;,\qquad B=1 \;, 
\label{eq:3.20}
\end{equation}
and to the free functions $A(\phi)$ and $V(\phi)$. In our 
notations this corresponds to identifying $A$ with $f$ and 
(\ref{eq:3.19}) with the Jordan frame action
\begin{equation}
S=\int d^{4}x\;\sqrt{-g}\left[\frac{f(\phi)}{16\pi 
G}R-\frac{1}{2}\nabla^{c}\phi\nabla_{c}\phi-V(\phi)\right]  
+S^{(m)}\left[g_{ab},\psi^{(m)}\right]\;.\label{eq:3.21}
\end{equation}
In fact, Flanagan's Jordan frame action can be generalized to 
arbitrary $B(\phi)$, which corresponds to our $\omega(\phi)$, 
obtaining the
Jordan frame action
\begin{equation}
S=\int d^{4}x\;\sqrt{-g}\left[ \frac{f(\phi)R}{16\pi 
G}-\frac{\omega(\phi)}{2}\nabla^{c} 
\phi\nabla_{c}\phi-V(\phi)\right] 
+S^{(m)}\left[g_{ab},\psi^{(m)}\right]\;.\label{eq:3.21b}
\end{equation}
$\bullet$~~~The \emph{Einstein frame with running units} 
corresponds to the  choice
\begin{equation}
A=1\;,\qquad B=1 \;,
\end{equation}
and to the free functions $\alpha(\phi)$ and $V(\phi)$. In our 
notations with a tilde denoting Einstein frame quantities, 
$\tilde{V}=\tilde{U}$
and $ e^{2\tilde{\alpha}}=\Omega^{-2}$ and the action 
(\ref{eq:3.19})
corresponds to
\begin{equation}
S= \int d^{4}x\;\sqrt{-g}\left\{ \frac{\tilde{R}}{16\pi 
G}-\frac{1}{2} \, \tilde{g}^{ab} \, \tilde{\nabla}_{a} 
\tilde{\phi}\tilde{\nabla}_{b}\tilde{\phi}-\tilde{U}( 
\tilde{\phi})+\Omega^{-2}\mathcal{L}^{(m)} 
\left[\tilde{g}_{ab},\psi^{(m)}\right]\right\} \;,\label{eq:3.23}
\end{equation}
$\bullet$~~~The \emph{Einstein frame with fixed units} 
corresponds to the 
choice
\begin{equation}
A=1\;,\qquad B=1\;,\qquad\alpha=0 \;,
\end{equation}
and, in our notations, to the action
\begin{equation}
S=\int d^{4}x\;\sqrt{-g} \left\{ \frac{\tilde{R}}{16\pi 
G}-\frac{1}{2}\tilde{g}^{cd} 
\tilde{\nabla}_{c} 
\tilde{\phi}\tilde{\nabla}_{d}\tilde{\phi}-\tilde{U}( 
\tilde{\phi})+\mathcal{L}^{(m)}\left[\tilde{g}_{ab}, 
\psi^{(m)}\right]\right\} \;,\label{eq:3.25}
\end{equation}
in which there is no anomalous coupling of the scalar $\tilde{\phi}$
to matter $\left(\alpha=0\right)$. The difference between Einstein
frame with running units and with fixed units is in the choice of
the function $\alpha$. It could be said that in the Einstein frame
with fixed units the function $\alpha$ is not correctly transformed
according to eq.~(\ref{eq:3add4}), while the functions $A$ and 
$B$
are transformed according to 
eqs.~(\ref{eq:3add5})-(\ref{eq:3add7}).
Keeping the units fixed in the Einstein frame causes the masses to
remain constant. In our first example of Sec.~3.1.2, this would 
correspond to replacing eq.~(\ref{eq:3.16}) with
\begin{equation}
\tilde{\square} 
\tilde{\psi}-\frac{\tilde{R}}{6}\tilde{\psi} 
-m^{2}\tilde{\psi}=0\;,\label{eq:3.26}
\end{equation}
with a constant mass $m$ introduced by hand. Of course, one can 
\emph{postulate} this equation, which is debatable, but it 
should at least be made clear that it can not be derived from 
eq.~(\ref{eq:3.14}) by  using Dicke's spacetime-dependent 
rescaling of units. In other words, the first two terms in 
eq.~(\ref{eq:3.26}) are obtained with a  conformal
transformation while the third one is arbitrarily replaced by 
$-m^{2}\tilde{\psi}$.

There are situations in cosmology in which the scalar field 
$\phi$ 
is assumed to dominate the dynamics of the universe and ordinary 
matter is ignored, setting $S^{(m)}=0$. In these situations 
(corresponding to $ m=0$ in the example of eqs.~(\ref{eq:3.14}) 
and  (\ref{eq:3.26})) the running of units does not matter as 
long as only cosmological dynamics is studied. However, the 
issue  will resurface whenever  massive test particles or test 
fields, or non-conformal matter are  introduced into this 
picture, or when redshift or reaction rates are considered 
\cite{Catenaetal}.

There is a way that is, in principle, consistent to obtain the 
Einstein frame with fixed units: if one introduces in the Jordan 
frame a factor that exactly compensates for the $\Omega^{-2}$ 
factor in front of the matter Lagrangian density when 
conformally transforming to the Einstein frame, the scalar 
$\tilde{\phi}$  will couple minimally  to matter in this frame. 
This is done, e.g., in  Ref.~\cite{Ferraris}.  However, the 
price to pay is the non-minimal coupling of $\phi$ to
matter, and the violation of the equivalence principle, in the Jordan
frame, which is alien from the spirit of Brans-Dicke and other 
scalar-tensor
theories.

\section{Energy conditions and singularity theorems}
\setcounter{equation}{0}

We now want to discuss the energy conditions in the Jordan and 
Einstein frame. Let us consider, for the sake of illustration, 
Brans-Dicke theory represented by the action
\begin{equation}
S=\int d^{4}x\; \sqrt{-g}\left[\phi 
R-\frac{\omega}{\phi}\nabla^{c}\phi\nabla_{c} 
\phi-V(\phi)\right]+S^{(m)} \label{eq:4.1}
\end{equation}
(the arguments proposed apply, however, to general scalar-tensor   
theories).
The field equations can be written as 
\begin{equation}
G_{ab}=\frac{8\pi}{\phi}T_{ab}^{(m)}+T_{ab}\left[\phi\right]\;,\label{eq:4.2}\end{equation}
\begin{equation}
\square\phi=\frac{1}{2\omega+3}\left[8\pi T^{(m)}+\phi\frac{dV }{d\phi}-2V\right]\;,\label{eq:4.3}\end{equation}
where \begin{equation}
T_{ab}\left[\phi\right]=\frac{\omega}{\phi^{2}}\left(\nabla_{a}\phi\nabla_{b}\phi-\frac{1}{2}g_{ab}\nabla^{c}\phi\nabla_{c}\phi\right)-\frac{V}{2\phi}g_{ab}+\frac{1}{\phi}\left(\nabla_{a}\nabla_{b}\phi-g_{ab}\square\phi\right)\;,\label{eq:4.4}\end{equation}
is often identified with an effective stress-energy tensor of the
scalar $\phi$. There are three possible ways of identifying an 
effective stress-energy tensor for $\phi$  
\cite{BellucciFaraoni,mybook} and the choice of 
eq.~(\ref{eq:4.4}) 
has sometimes been criticized in
the literature \cite{SantiagoSilbergleit,mybook,Torres}. If
the choice (\ref{eq:4.4}) is accepted, as is common in the literature,
it is easy to see that the strong, weak, and dominant energy conditions
of general relativity \cite{Wald} can all be violated by the scalar
$\phi$ regarded as an effective form of matter. This is due to the
non-canonical form of $T_{ab}$; the last term in 
eq.~(\ref{eq:4.4})
is linear in the second derivatives of $\phi$ instead of being quadratic
in the first derivatives, and it makes the sign of $T_{00}\left[\phi\right]$
indefinite, even causing negative energy densities. The possibility
of negative energy is regarded by certain authors as a criterion to
discard the Jordan frame \emph{a priori} as unphysical (see 
\cite{MagnanoSokolowski,FGN} for references). Since we 
know that Jordan frame and Einstein
frame with running units are physically equivalent, we argue that
these authors are left with the  ``Einstein frame with fixed 
units''
version of the theory, which is physically ill-motivated. Moreover,
it is not a negative kinetic energy that is worrisome, but rather
an energy that is unbounded from below, so that the system can decay
to lower and lower energy states {\em ad infinitum} (the 
electron in the
hydrogen atom has negative total energy but there is a ground state
of the Hamiltonian which corresponds to a minimum for the
spectrum of energy eigenvalues). Hence, the mere possibility of 
negative energies
is not, by itself, an argument to rule out the Jordan frame. As 
pointed
out in Ref.~\cite{Flanagan}, the energy conditions differ in the
two frames but there is no physical observable corresponding to the
sign of $G_{ab}u^{a}u^{b}$ for all timelike vectors $u^{a}$, hence
there is no measurable inconsistency between the two frames. 
Furthermore, a positive energy theorem has been shown to hold 
for special scalar-tensor theories in the Jordan frame 
\cite{Bertolamienergy}.

The validity of the energy conditions for the Einstein frame
scalar $\tilde{\phi}$ has been emphasized in relation with the Hawking-Penrose
singularity theorems \cite{HawkingEllis,Wald}. If the strong and
dominant energy conditions hold for $\tilde{\phi}$ and for ordinary
matter in the Einstein frame, the singularity theorems apply, even
though the same energy conditions are violated in the Jordan frame.
This situation is seen by some as the possibility to circumvent the
singularity theorems. In the cosmological  context this would 
imply
that it is possible to find solutions that are free of Big Bang
singularities just by going to the Jordan frame. This is clearly 
impossible
if these two conformal frames are physically equivalent: the absence
of singularities in one frame and their occurrence in the conformally
rescaled theory has thus lead to an apparent paradox 
\cite{529,725}. If, following Dicke \cite{Dicke}, the Jordan and 
Einstein frame are equivalent, singularities occur in the 
Einstein frame if and only if they occur
in the Jordan frame. The puzzle is quickly resolved as follows 
(see also Ref.~\cite{mybook}): consider the 
FLRW metric in the Jordan frame
\begin{equation}
ds^{2}=-dt^{2}+a^{2}(t)\left[\frac{dr^{2}}{1-Kr^{2}} 
+r^{2}\left(d\theta^{2}+\sin^{2}\theta 
\, d\varphi^{2}\right)\right]\;, 
\end{equation}
and its Einstein frame cousin
\begin{equation}
d\tilde{s}^{2} = -d\tilde{t}^{2}+ 
\tilde{a}^{2}(\tilde{t})\left[\frac{dr^{2}}{1 
-Kr^{2}}+r^{2}\left(d\theta^{2} +\sin^{2}\theta 
d\varphi^{2}\right)\right]\;, 
\end{equation}
with $ d\tilde{t}=\Omega \, dt$, $\tilde{a}=\Omega \, a$, and 
proper length
$ d\tilde{l}=\tilde{a} 
\, \left|d\underline{x}\right|=\Omega \, 
a\left|d\underline{x}\right|=\Omega \,  dl$. To ascertain 
whether there is a Big Bang or other singularity in the
Einstein frame with running units it is not  sufficient to 
examine the behavior of the scale factor $\tilde{a}(\tilde{t})$ 
as $\tilde{t}\rightarrow0$. One must instead study the ratio of 
a typical physical (proper) length
$\tilde{a}(\tilde{t})\left|d\underline{\tilde{x}}\right|$ to the
unit of length $\tilde{l}_{u}(\tilde{\phi})=\Omega \, l_{u}$, 
where
$l_{u}$ is the fixed length  unit in the Jordan frame and 
$\left| d\underline{\tilde{x}}\right|$
is the (comoving) coordinate distance in the Einstein frame. 
This ratio is
\begin{equation}
\frac{\tilde{a}( 
\tilde{t}) \left|d\underline{\tilde{x}}\right|}{\tilde{l}_{u}( 
\tilde{\phi})}=\frac{\Omega \, a(t)\left|d\underline{x} 
\right|}{\Omega \,
l_{u}}=\frac{a(t)\left|d\underline{x}\right|}{l_{u}} \;.
\end{equation}
Therefore, $\frac{ 
\tilde{a}\left|d\underline{\tilde{x}}\right|}{\tilde{l}_{u}( 
\tilde{\phi})}\rightarrow0$
if and only if $\frac{a\left|d\underline{x}\right|}{l_{u}} 
\rightarrow0$, or 
a singularity occurs in the Einstein frame if and only if it is 
present in the Jordan frame. The argument is not yet complete, 
because one has to make sure that the finite time at which the 
singularity occurs (``initial time'' for a Big Bang singularity) 
is not mapped  into an infinite time in the other frame. This is 
easily accomplished by examining the ratio of $\tilde{t}$ to 
the varying unit of time  $\tilde{t}_{u}(\tilde{\phi})=\Omega \, 
t_{u}$ in the Einstein frame, where $t_{u}$ is the fixed unit 
of  time in the Jordan frame. This ratio is
\begin{equation}
\frac{\tilde{t}}{\tilde{t}_{u}}=\frac{\int_{0}^{t} 
\Omega(\phi)dt'}{\Omega(\phi)t_{u}}\approx\frac{t}{t_{u}}\;,
\end{equation}
as $t\rightarrow0$ for an initial Big Bang singularity. 
Therefore, $\tilde{t}\rightarrow0$ in the Einstein frame is 
equivalent to $t\rightarrow0$ in the Jordan frame.

One can also check whether a singularity in the matter energy density
occurs in both frames. The energy density of the cosmic fluid transforms
as $\tilde{\rho}=\Omega^{-4}\rho$ on dimensional grounds (for a formal
derivation see, e.g.,  Ref.~\cite{mybook}). The Einstein frame 
unit of 
energy
is
\begin{equation}
\tilde{\rho}(\tilde{\phi}) 
\approx\frac{\tilde{m}}{\tilde{l}_{u}^{3}} 
=\frac{\Omega^{-1}m_{u}}{\Omega^{3}l_{u}^{3}} 
=\Omega^{-4}\rho_{u}\;, 
\end{equation}
where $\rho_{u}\approx  m / l_u^3 $ is the (constant) unit
of energy density in the Jordan frame. In a Big Bang singularity,
however, it is the ratio $ \tilde{\rho} / \tilde{\rho}_{u} $
that matters, not $\tilde{\rho}$. We have
\begin{equation}
\frac{\tilde{\rho}}{\tilde{\rho}_{u}}=\frac{\Omega^{-4}\rho}{\Omega^{-4}\rho_{u}}=\frac{\rho}{\rho_{u}}\;,
\end{equation}
hence $\frac{\tilde{\rho}}{\tilde{\rho}_{u}}\rightarrow\infty$ if
and only if $\frac{\rho}{\rho_{u}}\rightarrow\infty$, establishing
once again the equivalence of the two frames. If one were to consider
merely $\tilde{\rho}$ instead of $  \tilde{\rho} / 
\tilde{\rho}_{u} $, one would erroneously conclude that 
singularities occur in one conformal
frame but not in the other. This happens if the Einstein frame 
with fixed units is considered, which is not physically 
equivalent to the Jordan frame (if one wishes to regard it as a 
physical theory).

\section{The $\Lambda$ problem and the Cauchy problem}
\setcounter{equation}{0}

In this section we briefly discuss other issues in the realm of 
classical
physics in which the Jordan and the Einstein frame (with running units)
prove to be physically equivalent, in spite of claims to the 
contrary.
These are the cosmological constant problem and the Cauchy problem.

It has been claimed that the  issue of the conformal frame has 
implications for the notorious cosmological constant problem 
\cite{Weinberg} of why the cosmological constant energy 
density is 120 orders 
of magnitude smaller than what can be calculated with simple 
quantum mechanics. The stress-energy tensor  associated 
with a cosmological constant $ T_{ab}^{(\Lambda)}=\Lambda 
g_{ab}/\left( 8\pi G \right) $ provides a Jordan frame energy 
density 
$\rho_{\Lambda}= \Lambda / (8\pi G) $ which is constant, and a 
conformal cousin $\tilde{\rho}_{\Lambda}=\Omega^{-4}\rho_{\Lambda}=e^{-\alpha\tilde{\phi}}\Lambda$
in the Einstein frame, where $\alpha>0$ is an appropriate constant.
Thus, $\tilde{\rho}_{\Lambda}$ represents a decaying cosmological
``constant''; the opinion is often voiced that the exponential
factor $e^{-\alpha\tilde{\phi}}$ multiplying $\Lambda$ in the Einstein
frame helps alleviating, if not outright solving, the 
cosmological constant problem (see, e.g., Sec. 4.22 of 
Ref.~\cite{Fujii}).  Again, this would mean that the two 
conformal frames are physically  inequivalent
in contrast with the spirit of Dicke's paper 
\cite{footnote5}.

It is easy to  see that this argument fails to ease off the 
cosmological constant problem. Again, what matters in the 
Einstein frame is not
the form (or numerical value) of $\tilde{\rho}_{\Lambda}$, but the
ratio $ \tilde{\rho}_{\Lambda} / \tilde{\rho}_{u} $, where 
$\tilde{\rho}_{u}=\Omega^{-4}\rho_{u}$
is the unit of energy density in the Einstein frame, and $\rho_{u}$
is the corresponding Jordan frame unit. The ratio
\begin{equation}
\frac{\tilde{\rho}_{\Lambda}}{\tilde{\rho}_{u}}=\frac{\Lambda e^{-\alpha\tilde{\phi}}}{8\pi G\rho_{u}e^{-\alpha\tilde{\phi}}}=\frac{\Lambda}{8\pi G\rho_{u}}\;,
\end{equation}
is the same in the Jordan and Einstein frame and, barring unforeseen
complications at the quantum level, the cosmological constant problem
is not alleviated a bit by choosing the Einstein frame with 
running units.

Of course, one could then state that the Einstein frame with 
fixed units solves the problem because then one would consider 
$ \tilde{\rho}_{\Lambda}/ \rho_u \propto 
e^{-\alpha\tilde{\phi}}$
instead of $ 
\tilde{\rho}_{\Lambda} / \tilde{\rho}_u=$constant;
this would be nonsense because the cosmological constant cannot 
be calculated in one theory (where it is huge) and then mapped 
into the Einstein frame with fixed units which bears no physical 
relation with the original Jordan frame. $\Lambda$ should  
be calculated directly in the Einstein frame with fixed units 
and it is still huge. The argument presented here applies also 
to situations in which the cosmological ``constant'' term 
changes \cite{BertolamiNC}.

Finally, we want to comment on the Cauchy problem for  
scalar-tensor  gravity and its implications for the equivalence 
of the two conformal frames. The folklore about the Cauchy 
problem is that the mixing of the spin two and spin zero degrees 
of freedom $g_{ab}$ and $\phi$ in the Jordan frame makes these 
variables an inconvenient set for formulating the initial value 
problem, which is not well posed in the Jordan frame: on the 
other hand, the Einstein frame variables
$\left(\tilde{g}_{ab},\tilde{\phi}\right)$ admit a well-posed 
Cauchy problem completely similar to that of general relativity 
(see, e.g., the influential paper \cite{DamourFarese}). Were 
this true, it 
would  appear that the Jordan and Einstein frame are physically 
inequivalent in this respect. This position toward the Cauchy 
problem, however, ignores two older references showing that the 
Cauchy problem is  well posed {\em in the Jordan frame} for 
two  specific scalar-tensor theories: Brans-Dicke
theory with a free scalar $\phi$ \cite{CockeCohen} and the 
theory of a scalar field conformally coupled to the Ricci 
curvature \cite{Noakes}. The task of studying the Jordan 
frame Cauchy problem 
for \emph{general} scalar-tensor  theories has been taken on in 
a recent paper \cite{Salgado} in which it  is shown,
using generalized harmonic coordinates, that the Cauchy problem 
is well posed, although further study is necessary for 
implementing a full 3+1 formulation $\grave{a}$ la York 
\cite{York} in practical (numerical)
applications \cite{Salgado}. This shows that, contrary to the 
common lore, the Jordan and the Einstein frames are physically 
equivalent
also with respect to the initial value problem. The issue of mapping
the details of the Jordan frame Cauchy problem into details of the
corresponding Einstein frame problem and, in particular, clarifying
the role played by running units,  will be discussed elsewhere.
It is clear that  the equivalence between the two conformal 
frames breaks down when the conformal 
transformation breaks down, i.e., when $f(\phi)=0$ or $f_1\equiv 
2f+3\left( df/d\phi \right)^2=0$ (cf. eqs.~(\ref{eq:1.4}) and 
(\ref{eq:1.5})). However, the Jordan frame initial value problem 
may not be well posed as well when $f=0$, and requiring $f>0$ 
eliminates also the singularities $f_1=0$ (see 
Refs.~\cite{Abramoetal,PRDsingularities} for a discussion of 
these singularities and Ref.~\cite{Bronnikov} for conformal 
continuation past these points).

\section{Conclusions}

It appears that, at the classical level, the Jordan and Einstein 
frames are physically equivalent when the units of fundamental 
and derived quantities are allowed to scale appropriately  with 
the conformal factor $\Omega$ in the Einstein frame. Previous 
doubts on the physical equivalence with respect to the Cauchy 
problem \cite{DamourFarese} seem to dissipate in the light of 
recent work \cite{Salgado}, although a more comprehensive 
picture is 
desirable. The arguments against the equivalence of the two 
frames raised in the past regard positivity of the energy in the 
Einstein frame and the indefiniteness of its sign in the Jordan 
frame (\cite{MagnanoSokolowski,FGN} and references therein). 
This is particularly relevant at the quantum level: negative 
energies do not allow a stable ground state and 
the system would decay to a lower and lower energy states {\em 
ad infinitum}. However, as was pointed out in \cite{Flanagan}, 
there is no physical observable corresponding to the sign of 
$T_{ab}u^a u^b$ or $G_{ab}u^a u^b$, where $u^a$ is a timelike 
four-vector, and specific examples of scalar-tensor theories 
that are stable in the Jordan frame have been found 
\cite{Bertolamienergy}. The relevant question to ask, at least 
at the classical level, is not what the sign of the energy is, 
but rather whether the 
energy is bounded from below, which may well occur in 
scalar-tensor gravity (see, e.g., \cite{PRDenergy}).

At the quantum level, the issue of the ground state becomes more 
delicate, as there are more decay channels than at the 
classical level. Although the conformal equivalence seems to hold 
to some extent at the semiclassical level, in which the matter 
fields are quantized while the variables $\left( g_{ab}, \phi 
\right)$ are classical (see Ref.~\cite{Flanagan} for a brief 
discussion and references), this equivalence definitely breaks 
down when $\phi$ is quantized \cite{FGN}. When also $g_{ab} $ is 
quantized 
in full quantum gravity, inequivalent quantum theories have been 
found 
\cite{Flanagan,AshtekarCorichi,Elizaldeetal,Grumilleretal}. 
This is not surprising because the conformal transformation can 
be seen as a Legendre transformation \cite{MagnanoSokolowski}. A  
similar Legendre transformation is used in the classical 
mechanics of particles 
to switch from the canonical coordinates $q$ of the Lagrangian 
description to the variables $\left( q,p \right)$ of the 
Hamiltonian formalism, and this Legendre map is an example of a 
canonical transformation \cite{Goldsteinp385}. Now, it is well 
known that Hamiltonians that are classically equivalent become 
inequivalent when quantized: they exhibit different energy 
spectra and scattering amplitudes \cite{CalogeroDegasperis}. 
Therefore, we expect  the analogous ``canonical 
transformations'' between different conformal frames not to be 
unitary and to yield physically inequivalent  theories at the 
quantum  level \cite{footnote6}. A common objection to this 
statement arising among particle physicists is based on the 
equivalence theorem of Lagrangian field theory, which states 
that  the S-matrix is invariant under local (non-linear) field 
redefinitions \cite{equivalence}. Since the conformal 
transformation (\ref{eq:2.2}) and (\ref{eq:2.3}) is a local 
non-linear redefinition of the fields $g_{ab}$ and $\phi$ , it 
would seem that quantum physics is invariant under change of 
the conformal frame. However, this is not true in general 
because the field theory approach in which the equivalence 
theorem is derived applies to gravity only in the perturbative 
regime in which the fields describe small deviations from  
Minkowski space. In this regime, tree level quantities can be 
calculated in any conformal frame with the same results. 
However, when the metric tensor is allowed full dynamical 
freedom  and is not restricted to be a small perturbation of 
a fixed background, the field theory approach and the 
equivalence theorem do not apply. It is plausible that the 
equivalence theorem can be proved also for fixed backgrounds 
that are curved and do not coincide with the Minkowski space of 
effective field theory. However, we are not aware of such a  
generalization in the literature on quantum field theory on 
curved space (a proof of a generalized equivalence theorem will 
be pursued elsewhere). Thus, it is clear  that the equivalence 
theorem fails in the non-perturbative regime; nevertheless, one 
can consider semiclassical situations in which the metric is 
classical and the full scalar $\phi$ is quantized, and it is 
quite possible that the conformal transformation leaves the 
quantum physics of $\phi$ unaffected --- after all, the physics 
of the classical 
metric is invariant under change of conformal frame and 
quantization of a scalar field in a  fixed background metric 
poses no problems \cite{BirrellDavies}. Indeed, there are 
examples in which such semiclassical theories related by a 
conformal transformation seem to be equivalent 
\cite{Bertolamisemiclassical}. A precise 
and detailed understanding  of the conformal (in)equivalence at 
the quantum level, however, requires further work.

\section*{Acknowledgments}

We thank a referee for useful remarks and references. This work 
was supported by the Natural Sciences and Engineering  Research  
Council of Canada ({\em NSERC}).


\end{document}